\documentclass[12pt]{iopart}
\usepackage{iopams}
\usepackage{amssymb,latexsym}
\usepackage{amstext}

\begin{document}

\title[Dihedral symmetry of periodic chain]{Dihedral symmetry
 of periodic chain: \\ quantization and coherent states}
 \author{P Luft, G Chadzitaskos and  J Tolar}
\address{Department of Physics\\
Faculty of Nuclear Sciences and Physical Engineering         \\
Czech Technical University \\ B\v rehov\'a 7,  CZ - 115 19 Prague,
Czech Republic}
 \ead{jiri.tolar@fjfi.cvut.cz}

 \begin{abstract}
 Our previous work on quantum kinematics and coherent states over
finite configuration spaces is extended: the configuration space is,
as before, the cyclic group $\mathbf{Z_{n}}$ of arbitrary order
$n=2,3,\ldots$, but a larger group --- the non-Abelian dihedral
group $\mathbf{D_{n}}$ --- is taken as its symmetry group. The
corresponding group related coherent states are constructed and
their overcompleteness proved. Our approach based on geometric
symmetry can be used as a kinematic framework for matrix methods in
quantum chemistry of ring molecules.
\end{abstract}

 \submitto{J. Phys. A: Math. Theor.}
%\maketitle


\section{Introduction}

The mathematical arena for ordinary quantum mechanics is, due to
Heisenberg's commutation relations, the infinite-dimensional Hilbert
space. A useful model for quantum mechanics in a Hilbert space of
finite dimension $n$ is due to H. Weyl \cite{Weyl}. Its geometric
interpretation as the simplest quantum kinematic on a finite
discrete configuration space formed by a periodic chain of $n$
points, was elaborated by J.~Schwinger \cite{Schwinger}. In
\cite{Tolar, StovTolar} we proposed a group theoretical formulation
of this quantum model in terms of Mackey's quantization
\cite{Mackey, HDDTolar}. It is based on Mackey's system of
imprimitivity which represents a group theoretical generalization of
Heisenberg's commutation relations.

The geometrical picture behind the group theoretical approach is the
following \cite{HDDStovTolar}: one has a discrete or continuous
configuration space together with a geometrical symmetry group
acting transitively on it, i.e. the configuration space is a
homogeneous space of the group. In particular, Weyl's model is based
on configuration space $\mathbf{Z_{n}}$ (where $\mathbf{Z_{n}}$ is
the cyclic group of order $n=2,3,\ldots $) with symmetry
$\mathbf{Z_{n}}$ acting on the periodic chain $\mathbf{Z_{n}}$ by
discrete translations. In this paper our formulation of Weyl's model
is generalized by extending the Abelian symmetry group
$\mathbf{Z_{n}}$ of the periodic chain to the dihedral group
$\mathbf{D_{n}}$ --- the non-Abelian symmetry group of a regular
$n$-sided polygon.

Coherent states belong to the most important tools in many
applications of quantum physics. They found numerous applications in
quantum optics, quantum field theory, condensed matter physics,
atomic physics etc. There are various definitions and approaches to
the coherent states dependent on author and application. Our main
reference is \cite{Perelomov}, where the systems of coherent states
related to Lie groups are described. The basic feature of such
systems is that they are overcomplete. As shown for instance in
\cite{TCh}, Perelomov's method can be equally well applied to
discrete groups. Starting with irreducible systems of imprimitivity
we shall construct irreducible sets of generalized Weyl operators,
whose action on properly chosen vacuum states will produce the
resulting families of coherent states.

In section 2 after recalling Mackey's Imprimitivity Theorem for
finite groups \cite{Coleman} the construction of systems of
imprimitivity is described. Then necessary notations for the
dihedral groups are introduced in section 3. Section 4 is devoted to
the construction of the two irreducible systems of imprimitivity for
$\mathbf{D_{n}}$ based on $\mathbf{Z_{n}}$, each consisting of a
projection--valued measure and an induced unitary representation.
From them, the corresponding quantum position and momentum
observables are constructed in section 5. This is the starting point
for construction of the set of generalized Weyl operators and
generalized coherent states in section 6. We apply the method of
paper \cite{TCh}, where quantization on $\mathbf{Z}_{n}$ with
Abelian symmetry group $\mathbf{Z}_{n}$ and the corresponding
coherent states were investigated. Concluding section 7 contains
remarks concerning the replacement of the Abelian cyclic symmetry
group $\mathbf{Z}_{n}$ by the non-Abelian dihedral group
$\mathbf{D}_{n}$ as the group of motions of the configuration space
$\mathbf{Z}_{n}$. The interesting feature of our construction is the
fact that, even if the group property of the set of Weyl operators
is lost, the families of coherent states still possess the required
overcompleteness property.

\section{Systems of imprimitivity for finite groups}

We consider the case when the configuration space $\mathbf{M}$ and
its symmetry group $\mathbf{G}$ are finite. Our configuration space
will be a finite set $\mathbf{M}= \{m_{1}, m_{2},...,m_{n}\}$,
$n=|\mathbf{M}|$. Let $\mathbf{G}$ be a finite group acting
transitively on $\mathbf{M}$, and let $\mathbf{H}$ be the stability
subgroup. Let $\mathbf{L}$ be an irreducible unitary representation
of subgroup $\mathbf{H}$ on Hilbert space
$\mathcal{H}^{\mathbf{L}}$.

System of imprimitivity is a pair $(\mathbf{V},\mathbf{E})$, where
$\mathbf{E}$ is a projection-valued measure on configuration space
$\mathbf{G}/\mathbf{H}$ and $\mathbf{V}$ is a unitary representation
of the symmetry group $\mathbf{G}$ such that
\begin{equation}
\mathbf{V}(g)\mathbf{E}(S)\mathbf{V}(g)^{-1}=\mathbf{E}(g.S) \quad
\text{for all} \quad g\in\mathbf{G}, S \subset
\mathbf{G}/\mathbf{H}.
\end{equation}
In a finite-dimensional Hilbert space $\mathcal{H}= \mathbb{C}^{n}$
the standard projection-valued measure is given by finite sums of
diagonal matrices
\begin{equation}
  \mathbf{E}(m_{i}) := \text{diag}(0,0,...,1,...,0), \; i =1,2,...,n.
\end{equation}

The Imprimitivity Theorem for finite groups has the following form
\cite{Coleman}: \\
 {\bf Theorem :}  {\it A unitary representation
$\mathbf{V}$ of a finite group $\mathbf{G}$ in Hilbert space
$\mathcal{H}$ belongs to the imprimitivity system
$(\mathbf{V},\mathbf{E})$ with standard projection-valued measure
based on $\mathbf{G}/\mathbf{H}$, if and only if $\mathbf{V}$ is
equivalent to an induced representation
$Ind_{\mathbf{H}}^{\mathbf{G}}(\mathbf{L})$ for some unitary
representation $\mathbf{L}$ of subgroup $\mathbf{H}$. The system of
imprimitivity is irreducible, if and only if $\mathbf{L}$ is
irreducible.}

Thus a unitary representation $\mathbf{V}$ for a system of
imprimitivity is constructed directly as an induced representation.
Let $\mathbf{G}$ be a finite group of order $r$, $\mathbf{H}$ its
subgroup of order $s$. Suppose that $\mathbf{L}$ is a representation
of the subgroup $\mathbf{H}$. Let us decompose the group
$\mathbf{G}$ into left cosets
\begin{equation}\label{cosets}
  \mathbf{G}=\{\bigcup_{j=1}^{r/s}t_{j}\cdot\mathbf{H}
  \;  \vert \; t_{j} \in \mathbf{G}, \; t_{1}=e  \}.
\end{equation}
Group elements $t_{j}$ are arbitrarily chosen representatives of
left cosets. If the dimension of the representation $\mathbf{L}$ is
$l$, then the induced representation $\mathbf{V}$ of $\mathbf{G}$ is
given by
%\begin{equation}
%  (\mathbf{V}(g))_{ij}:=
%  \begin{cases}
%    \mathbf{L}(h)  &  \rm{if} \quad t^{-1}_{i}\cdot g \cdot t_{j} = h \;
%    \rm{for some} \; h \in \mathbf{H}, \\
%    0  &  \rm{otherwise};
%  \end{cases}
%\end{equation}
\begin{eqnarray}
(\mathbf{V}(g))_{ij}& = & \mathbf{L}(h) \quad \text{ if } \quad
t^{-1}_{i}\cdot g \cdot t_{j} = h \quad \text{for} \; \mathrm{some}
\;
h \in \mathbf{H}, \\
& = & 0 \quad  \text{ otherwise };
\end{eqnarray}
here $(\mathbf{V}(g))_{ij}$ are $l \times l$ matrices which serve as
building blocks for
\begin{equation}
 \mathbf{V}(g) = Ind_{\mathbf{H}}^{\mathbf{G}}(\mathbf{L})
\end{equation}
and the subscript $ij$ denotes the position of the block in
$\mathbf{V}(g)$.

\section{Structure of dihedral groups}

The dihedral group $\mathbf{D_{n}}$, where $n=2,3,\ldots$, is a
non-Abelian finite group of order $2n$ with the structure of a
semidirect product of two cyclic groups:
\begin{equation}\label{semi}
  \mathbf{D_{n}} = \mathbf{Z_{n}} \triangleright \mathbf{Z_{2}}.
\end{equation}
It arises as the symmetry group of a regular polygon and is
generated by discrete rotations and reflections. The elements of the
subgroups $\mathbf{Z_{2}}$ and $\mathbf{Z_{n}}$ will be denoted
\begin{equation}\label{ZN}
  \mathbf{Z_{2}} = \{+1,-1\}; \quad \mathbf{Z_{n}} =
  \{e=r_{0},r_{1},...,r_{n-1}\}.
\end{equation}
 Group operation in $\mathbf{Z_{2}}$ is multiplication,
in $\mathbf{Z_{n}}$  $r_{i}\cdot r_{j}=r_{i+j \pmod{n}}$.

The multiplication law of the semidirect product \eref{semi} is
determined by a fixed homomorphism $f$ from $\mathbf{Z_{2}}$ to the
group of all automorphisms of the group $\mathbf{Z_{n}}$, $
f:\mathbf{Z_{2}} \rightarrow Aut(\mathbf{Z_{n}})$:
\begin{equation}\label{multlaw}
  (r_{i},x)\cdot (r_{j},y) = (r_{i}\cdot f(x)(r_{j}),x\cdot y), \;
x,y \in  \mathbf{Z_{2}}, \; r_{i},r_{j} \in \mathbf{Z_{n}}.
\end{equation}
Under this multiplication law, $\mathbf{Z_{n}}$ is a normal
subgroup. Specifically for $\mathbf{D_{n}}$, the mapping $f$ is
simply
\begin{equation}\label{}
  f:+1 \mapsto Id, \qquad f:-1 \mapsto Inv,
\end{equation}
where $Id$ is the identical mapping on $\mathbf{Z_{n}}$, $Inv$ is an
automorphism of $\mathbf{Z_{n}}$ which maps an element of
$\mathbf{Z_{n}}$ into its inverse:
\begin{equation}\label{}
  Inv: r_{k}\mapsto r_{k}^{-1}=r_{-k \pmod{n}}, \qquad r_i \in \mathbf{Z_{n}}.
\end{equation}
We shall need the explicit form of the multiplication law:
\begin{equation}\label{nasa}
  (r_{i},+1)\cdot (r_{j},x) = (r_{i}\cdot r_{j},x) =
  (r_{i+j \pmod{n}},x),
\end{equation}
\begin{equation}\label{nasb}
 (r_{i},-1)\cdot (r_{j},x) = (r_{i}\cdot r_{j}^{-1},-x) =
 (r_{i-j \pmod{n}},-x).
\end{equation}
\medskip

Thus the elements of $\mathbf{D_{n}}$ can be divided  in two
disjoint subsets:
\begin{enumerate}
\item The subset $ \{(r_{k},+1), \, k = 0,1,...,n-1 \}$ forms the subgroup
isomorphic to $\mathbf{Z_{n}}$ and the elements $(r_{k},+1)$ have
the geometrical meaning of integral multiples of a clockwise
rotation of an $n$-sided regular polygon through an angle $2\pi /n$.
\item The subset $ \{(r_{k},-1), \, k = 0,1,...,n-1 \}$ consists of
mirror symmetries with respect to axes in the $n$--sided polygon: if
$n$ is odd, then all axes of mirror symmetries pass through vertices
of the $n$--sided polygon; if $n$ is even, then only one half of
mirror symmetries have axes passing through opposite vertices, the
remaining axes are symmetry axes of two opposite sides of the
polygon.
\end{enumerate}

Summarizing, the group $\mathbf{D_{n}}$ consists of $n$ rotation
symmetries $\mathbf{R}_{k} = (r_{k},+1)$ and $n$ mirror symmetries
$\mathbf{M}_{k} = (r_{k},-1)$ obeying the following multiplication
rules (with $i,j = 0,1,...,n-1$):
\begin{equation}\label{nas1}
  \mathbf{R}_{i}\cdot\mathbf{R}_{j} = \mathbf{R}_{i+j \pmod{n}},
  \qquad
  \mathbf{R}_{i}\cdot\mathbf{M}_{j} = \mathbf{M}_{i+j \pmod{n}},
\end{equation}
\begin{equation}\label{nas2}
  \mathbf{M}_{i}\cdot\mathbf{R}_{j} = \mathbf{M}_{i-j \pmod{n}},
   \qquad
  \mathbf{M}_{i}\cdot\mathbf{M}_{j} = \mathbf{R}_{i-j \pmod{n}}.
\end{equation}

\section{Quantization on $\mathbf{Z_{n}}$ with $\mathbf{D_{n}}$ as
a symmetry group}

The configuration space $\mathbf{Z_{n}}$ will be identified with the
set of vertices of a regular $n$--sided polygon. We have seen that
$\mathbf{D_{n}}$ acts on $\mathbf{Z_{n}}$ transitively as a group of
discrete rotations and mirror symmetries. The stability subgroup
$\mathbf{H_{n}}$ of $\mathbf{D_{n}}$ is $\mathbf{Z_{2}}$ for all
$n$, hence we can write $ \mathbf{Z_{n}} \cong
\mathbf{D_{n}}/\mathbf{Z_{2}}$.

The stability subgroup $\mathbf{Z_{2}}$ is independent of the order
of symmetry group $\mathbf{D_{n}}$ and it has exactly two
inequivalent irreducible unitary representations, the trivial
representation
\begin{equation}\label{}
  \mathbf{T_{1}}:\mathbf{Z_{2}}\rightarrow \mathbb{C}:\pm 1 \mapsto 1,
\end{equation}
and the alternating representation
\begin{equation}\label{}
  \mathbf{T_{2}}:\mathbf{Z_{2}}\rightarrow \mathbb{C}:+1 \mapsto +1,
  \quad -1 \mapsto -1.
\end{equation}

Now the inequivalent quantum kinematics on the configuration space
$\mathbf{Z_{n}}$ are determined by inequivalent systems of
imprimitivity on $\mathbf{Z_{n}}$ with the symmetry group
$\mathbf{D_{n}}$. We require irreducibility of systems of
imprimitivity in order that the corresponding kinematical
observables act irreducibly in the Hilbert space. There will be
exactly two inequivalent irreducible systems of imprimitivity
$(\mathbf{V_1},\mathbf{E_1})$ and $(\mathbf{V_2},\mathbf{E_2})$ with
representations induced from irreducible unitary representations
$\mathbf{T_{1}}$ and $\mathbf{T_{2}}$.

 In both cases the Hilbert space $\mathcal H$ of quantum mechanics is the
space of complex functions on the configuration space
$\mathbf{Z_{n}}$ and it is isomorphic to $n$--dimensional complex
vector space $\mathbb{C}^{n}$ with standard inner product
\begin{equation}\label{}
<z_1,z_2>=\sum_{i=0}^{n-1}\bar{z}_{1i}z_{2i}.
\end{equation}
The standard projection-valued measure $\mathbf{E}$ is common to
both systems of imprimiti\-vi\-ty $(\mathbf{V_1},\mathbf{E})$ and
$(\mathbf{V_2},\mathbf{E})$. It is diagonal and generated by sums of
one-dimensional orthogonal projectors on $\mathbb{C}^{n}$ of the
form
\begin{equation}\label{proj1}
 \mathbf{E}(r_{i}) =  {}_{i}\left(
 \begin{array}{ccccc}
  {} & {} &  {}^{i} & {} & {} \, {}  \\
  {} & {} & \cdot & {} & {} \, {} \\
    \cdot & \cdot & 1 & \cdot & \cdot \, {} \\
  {} & {} & \cdot & {} & {} \, {}\\
  {} & {} & \cdot & {} & {} \, {}
  \end{array}\right), \quad i=0,1,...,n-1;
\end{equation}
Measure of an empty set in $\mathbf{Z_{n}}$ is the vanishing operator
on $\mathbb{C}^{n}$, measure of the whole configuration space is the
unit operator.

In order to obtain the two irreducible systems of imprimitivity, we
shall construct the representations induced from $\mathbf{T_{1}}$
and $\mathbf{T_{2}}$ on $\mathbb{C}^{n}$,
\begin{equation}\label{}
  \mathbf{V_{1}} =
  Ind_{\mathbf{Z_{2}}}^{\mathbf{D_{n}}}(\mathbf{T_{1}}), \qquad
  \mathbf{V_{2}} =
  Ind_{\mathbf{Z_{2}}}^{\mathbf{D_{n}}}(\mathbf{T_{2}}).
\end{equation}
According to \eref{cosets} the symmetry group $\mathbf{D_{n}}$ is
decomposed into left cosets,
\begin{equation}\label{}
  \mathbf{D_{n}}=
  \{\bigcup_{m=0}^{n-1}t_{m}\cdot\mathbf{Z_{2}}|t_{m}\in \mathbf{D_{n}}, \; t_{0} =
  e.\}
\end{equation}
In our case we have $\mathbf{Z_{2}} = \{\mathbf{R}_{0},\,
\mathbf{M}_{0}\}$; with the choice of coset representatives $t_{m} =
\mathbf{R}_{m}, \; m=0,1,...,n-1$, we obtain the decomposition
\begin{equation}\label{}
  \mathbf{D_{n}} = \{ \{\mathbf{R}_{0}, \mathbf{M}_{0}\} \cup \,
  \{\mathbf{R}_{1}, \mathbf{M}_{1}\} \cup
  \,... \cup \{\mathbf{R}_{n-1}, \mathbf{M}_{n-1}\} \}.
\end{equation}

Matrices of induced representations are then constructed in block
form: dimensions of both representations $\mathbf{V_{1}}$ and
$\mathbf{V_{2}}$ are equal to $n$,
\begin{equation}\label{}
  \text{dim}(\mathbf{V_{l}}) =
  \frac{|\mathbf{D_{n}}|}{\mathbf{|Z_{2}|}}\cdot
  \text{dim}(\mathbf{T_{l}})=\frac{2n}{2}\cdot 1 = n, \quad l=1,2,
\end{equation}
and matrix elements ($1\times 1$--blocks) have the following form:
  \begin{eqnarray}\label{ind1}
\mathbf{V_{l}}(g)_{ij}&=&
    \mathbf{T_{l}}(h) \quad  \text{if}\quad
    t^{-1}_{i}\cdot g \cdot t_{j} = h \quad
    \text{for some}\quad h \in \mathbf{Z_{2}}, \cr
    &=& 0 \quad \text{otherwise}.
  \end{eqnarray}
In our case $t_{i} = \mathbf{R}_{i}$, so the matrix element
$(\mathbf{V_{i}}(g))_{ij}$ does not vanish if and only if
\begin{equation}\label{ind2}
\mathbf{R}_{-i \pmod{n}}\cdot g \cdot \mathbf{R}_{j} \in
\{\mathbf{R}_{0}, \, \mathbf{M}_{0}\}.
\end{equation}

To construct the induced representation $\mathbf{V_{1}}$ --- first
for the subgroup of discrete rotations $g=\mathbf{R}_{k}$ ---
condition \eref{ind2}
\begin{equation}\label{}
 \mathbf{R}_{-i \pmod{n}}\cdot \mathbf{R}_{k} \cdot \mathbf{R}_{j}
  = \mathbf{R}_{-i+j+k  \pmod{n}}\in \{\mathbf{R}_{0}, \, \mathbf{M}_{0}\}
  \end{equation}
is equivalent to $i=j+k {\pmod{n}}$, hence matrix elements
\eref{ind1} of discrete rotations are
\begin{equation}\label{v1op}
  \ (\mathbf{V_{1}}(\mathbf{R}_{k}))_{ij}=\delta_{i,j+k \pmod{n}}.
\end{equation}
So the entire matrix is
\begin{equation}\label{v1rk}
  \mathbf{V_{1}}(\mathbf{R}_{k})=_{k}
  \left(\begin{array}{cccccccccc}
    &  &  &  & ^{k} & 1 &  &  &  &  \\
     &  &  &  &  &  & 1 &  &  &  \\
     &  &  &  &  &  &  & \cdot &  &  \\
     &  &  &  &  &  &  &  & \cdot &  \\
     &  &  &  &  &  &  &  &  & 1 \\
    1 &  &  &  &  &  &  &  &  &  \\
     & 1 &  &  &  &  &  &  &  &  \\
     &  & \cdot &  &  &  &  &  &  &  \\
     &  &  & \cdot &  &  &  &  &  &  \\
     &  &  &  & 1 &  &  &  &  &  \
  \end{array}\right).
\end{equation}

For the representation $\mathbf{V_{1}}$ of mirror symmetries
$g=\mathbf{M}_{k}$ condition (\ref{ind2}) acquires the form
\begin{equation}\label{}
  \mathbf{R}_{-i \pmod{n}}\cdot \mathbf{M}_{k} \cdot
  \mathbf{R}_{j} = \mathbf{M}_{-i-j+k \pmod{n}}
\in \{\mathbf{R}_{0}, \, \mathbf{M}_{0}\} \Leftrightarrow i = k-j
\end{equation}
due to (\ref{nas1}) - (\ref{nas2}), so the matrix elements
(\ref{ind1}) of mirror symmetries are
\begin{equation}\label{}
  (\mathbf{V_{1}}(\mathbf{M}_{k}))_{ij} = \delta_{i,k-j \pmod{n}}.
\end{equation}
The matrix $\mathbf{V_{1}}(\mathbf{M}_{k})$ has the explicit form
\begin{equation}\label{v1mk}
  \mathbf{V_{1}}(\mathbf{M}_{k}) =
  \left(\begin{array}{ccccccccc}
     &  &  &  & 1 &  &  &  &  \\
     &  &  & \cdot &  &  &  &  &  \\
    _{k} &  & 1 &  &  &  &  &  &  \\
     & \cdot &  &  &  &  &  &  &  \\
     1 &  & &  &  &  &  &  &  \\
     &  &  &  &  &  &  &  & 1 \\
     &  &  &  &  &  &  & \cdot &  \\
     &  &  &  &  &  & \cdot &  &  \\
     &  &  &  &  & 1 &  &  &  \
  \end{array}\right)
\end{equation}

The second representation $\mathbf{V_{2}}$ is obtained similarly via
(\ref{ind1}) as the representation induced from $\mathbf{T_{2}}$
with the result
\begin{equation}\label{jd}
  \mathbf{V_{2}}(\mathbf{R}_{k}) = \mathbf{V_{1}}(\mathbf{R}_{k}),
  \quad \mathbf{V_{2}}(\mathbf{M}_{k}) =
  -\mathbf{V_{1}}(\mathbf{M}_{k}).
\end{equation}
The representations $\mathbf{V_{1}}$ and $\mathbf{V_{2}}$ are
unitary, reducible and inequivalent; as could be expected,
the two systems of imprimitivity
differ only on reflections in $\mathbf{D_{n}}$.

\section{Quantum observables}

The basic quantum observables --- position and momentum operators
--- defining quantum kinematics on a configuration space have
natural definition if a system of imprimitivity is given.

 Classical position observable is a Borel mapping from the
configuration space, in our case from $\mathbf{Z_{n}}$, to the set
of real numbers. For the classical position observable counting
the points in $\mathbf{Z_{n}}$,
\begin{equation}\label{}
  f:\mathbf{Z_{n}}\rightarrow \mathbb{R}:r_{k}\mapsto k, \quad
  k=0,1,...,n-1,
\end{equation}
the corresponding {\it quantized position operator}
$\widehat{\mathbf{Q}}$ is expressed in terms of the
projection-valued measure (\ref{proj1}) as follows
\cite{HDDStovTolar}:
\begin{equation}\label{posop}
  \widehat{\mathbf{Q}}:=\sum_{k=0}^{n-1}k\cdot\mathbf{E}(f^{-1}(k))
  =\sum_{k=0}^{n-1}k\cdot\mathbf{E}(r_{k})
 = \text{diag}(0,1,\ldots,n-1).
\end{equation}
%i.e.
%\begin{equation}\label{posop}
% \widehat{\mathbf{Q}}=\begin{pmatrix}
%   0 &  &  &  &  \\
%     & 1 &  &  &  \\
%    & & \cdot &  &  \\
%    &  &  & \cdot &  \\
%    &  &  &  & n-1 \
% \end{pmatrix}
%\end{equation}
Note that the position operator is the same for both
systems of imprimitivity constructed in
previous section, i.e. in both quantum kinematics.

In the continuous case, {\it quantized momentum operators}
are obtained from unitary representation $\mathbf{V}$
by means of Stone's theorem \cite{BEH}:
{\it To each one-parameter subgroup $\gamma(t)$ of a symmetry group
there exists a self-adjoint operator
 $\widehat{\mathbf{P}}$ such that}
\begin{equation}\label{}
  \mathbf{V}(\gamma(t))=\exp(-it\widehat{\mathbf{P}}), \;\; t \in
  \mathbb{R}.
\end{equation}
However, this is not possible in the discrete case. One has to look
for self--adjoint operators $\widehat{\mathbf{P_l}}_{g}$ on
$\mathbb{C}^{n}$ such that
\begin{equation}\label{exp}
 \mathbf{V_{l}}(g)=\exp(-i\widehat{\mathbf{P_l}}_{g}),
 \qquad l=1,2, \quad g \in \mathbf{D_{n}}.
\end{equation}
One may try to compute the operators
$\widehat{\mathbf{P_l}}_{g}$ by inverting the exponential
(\ref{exp}),
\begin{equation}\label{ln}
  \widehat{\mathbf{P_l}}_{g}= i\cdot \ln(\mathbf{V_{l}}(g)),
\end{equation}
but then has to face the problem that the complex exponential is not
invertible, so the operators $\widehat{\mathbf{P_l}}_{g}$ will not
be determined uniquely.

Computation of functions of matrices is possible via the
Lagrange--Sylvester theorem (see the Appendix).
However, the spectral data needed there have their own physical
importance in quantum mechanics, so they will be
determined below for the operators $\mathbf{V_{1}}(\mathbf{R_{k}})$ and
$\mathbf{V_{1}}(\mathbf{M}_{k})$, $k=0,1,...,n-1$.
Because of (\ref{jd}) they are applicable to the other
system of imprimitivity, too.

Let us start with discrete rotations.
The eigenvalues of operator $\mathbf{V_{1}}(\mathbf{R_{1}})$ are
solutions of the secular equation
\begin{equation}\label{det}
  \det(\lambda\mathbb{I}-\mathbf{V_{1}}(\mathbf{R_{1}}))=0
\qquad  \text{or} \qquad \lambda^{n}-1=0,
\end{equation}
hence the spectrum is
\begin{equation}\label{}
  \sigma(\mathbf{V_{1}}(\mathbf{R}_{1}))=\{\lambda_{j}=e^{\frac{2\pi
  ij}{n}}|j=0,1,...,n-1\}.
\end{equation}
Then the eigenvalues of operators $\mathbf{V_{1}}(\mathbf{R_{k}})$ are
simply the powers of those of $\mathbf{V_{1}}(\mathbf{R_{1}})$,
\begin{equation}\label{spek1}
 \sigma(\mathbf{V_{1}}(\mathbf{R}_{k}))=
 \sigma(\mathbf{V_{1}}((\mathbf{R}_{1})^{k}))=
\{\lambda_{j}^k=e^{\frac{2\pi ijk}{n}}|j=0,1,...,n-1\}.
\end{equation}
Similarly the spectra of operators $\mathbf{V_{1}}(\mathbf{M}_{k})$
for mirror symmetries are obtained by solving
\begin{equation}\label{det1}
  \det(\lambda\mathbb{I}-\mathbf{V_{1}}(\mathbf{M}_{k}))=0,
\end{equation}
but here two cases should be distinguished.
\begin{enumerate}
   \item If $n$ is {\it odd}, then
(\ref{det1}) becomes
\begin{equation}\label{spm}
  (1-\lambda)(\lambda^{2}-1)^{\frac{n-1}{2}}=0
\qquad  \Rightarrow  \qquad
 \sigma(\mathbf{V_{1}}(\mathbf{M}_{k})) = \{+1,-1\}
\end{equation}
and the multiplicities of eigenvalues $\pm 1$ are $\frac{n\pm
1}{2}$.
    \item If $n$ is {\it even}, then the
characteristic polynomial of operator
$\mathbf{V_{1}}(\mathbf{M}_{k})$ depends, in addition to dimension
$n$, also on parameter $k$. At this point we have also to
distinguish if $k$ is odd or even. In the geometric picture
 we have to distinguish if the axis of mirror
symmetry $\mathbf{M}_{k}$ passes through opposite vertices of the
$n$--sided regular polygon ($k$ even), or if it is an axis of two
opposite sides of the polygon ($k$ odd). So if $n$ is even, then
(\ref{det1}) has following form:
\begin{eqnarray}\label{}
  0&=&
    (1-\lambda)^{\frac{n}{2}+1}(1+\lambda)^{\frac{n}{2}-1}
  \text{ if  $k$  is even } , \\
  0&=&  (1-\lambda)^{\frac{n}{2}}(1+\lambda)^{\frac{n}{2}}
  \text{ if $k$  is odd }.
\end{eqnarray}
The spectra for both cases are the same as for odd $n$, but the
multiplicities of eigenvalues are different. If $k$ is even, the
multiplicity of eigenvalue $+1$ is $\frac{n}{2}+1$, the multiplicity
of eigenvalue $-1$ is $\frac{n}{2}-1$; if $k$ is odd, then the
multiplicity of both eigenvalues is $\frac{n}{2}$.
\end{enumerate}

The evaluation of operators
$\widehat{\mathbf{P_1}}_{\mathbf{R}_{k}}$ for discrete rotations can
be done using the fact that rotations $\mathbf{R}_{k}$ form an
Abelian subgroup $\mathbf{Z_{n}}$ of $\mathbf{D_{n}}$. Thus we have
simply
\begin{equation}\label{}
  \exp(-i\widehat{\mathbf{P}}_{\mathbf{R}_{k}})=
  \mathbf{V_{1}}(\mathbf{R}_{k})=(\mathbf{V_{1}}(\mathbf{R}_{1}))^{k}
  = \exp(-ik\widehat{\mathbf{P}})
\end{equation}
where
$ \widehat{\mathbf{P}}=\widehat{\mathbf{P_1}}_{\mathbf{R}_{1}}$
 can be interpreted as self--adjoint momentum operator.
The spectrum (\ref{spek1}) of $\mathbf{V_{1}}(\mathbf{R}_{1})$
has $n$ different simple eigenvalues
$\lambda_{k}=e^{\frac{2\pi ik}{n}}$, so it remains to find
the corresponding one--dimensional spectral projectors
\begin{equation}\label{proj2}
  \mathbb{P}_{k}=|k\rangle \langle k|.
\end{equation}
Here $|k\rangle$ are  normalized eigenvectors of operator
$\mathbf{V_{1}}(\mathbf{R}_{1})$ belonging to eigenvalues
$\lambda_{k}$ \cite{StovTolar}:
\begin{equation}\label{}
  |k\rangle = \frac{1}{\sqrt{n}}\left(\begin{array}{c}
    \lambda_{k}^{n-1} \\
    \lambda_{k}^{n-2} \\
    \cdot \\
    \cdot \\
    \lambda_{k} \\
    1
  \end{array}\right),
\end{equation}
Using (\ref{proj2}), matrix elements of $\mathbb{P}_{k}$ can be written as
\begin{equation}\label{}
  (\mathbb{P}_{k})_{lm}=
  \frac{1}{n}\lambda_{k}^{n-l}\overline{\lambda_{k}^{n-m}}=
  \frac{1}{n}e^{\frac{2\pi ik(m-l)}{n}}.
\end{equation}
Then, using (\ref{ln}) for simple eigenvalues, we have
\begin{equation}\label{}
 (\widehat{\mathbf{P}})_{lm}=
  i (\ln \mathbf{V_{1}}(\mathbf{R}_{1}))_{lm}=
 = i \sum_{j=0}^{n-1}\ln (\lambda_j)(\mathbb{P}_{j})_{lm},
\end{equation}
hence matrix elements of the momentum operator are obtained:
\begin{eqnarray}\label{gr}
 (\widehat{\mathbf{P}})_{lm}
  &=&  \frac{2\pi}{n}\frac{1}{1-e^{\frac{2\pi i(m-l)}{n}}} \quad m \neq l, \\
  &=&  -\pi\frac{n-1}{n} \quad m = l.
\end{eqnarray}
Note that this result was obtained in \cite{TCh} by finite Fourier
transform of the position operator. For the analysis of
operators of mirror symmetries see the Appendix.
From the physical point of view unitary operators
$\mathbf{V_{1,2}}(\mathbf{M}_{k})$ play the role of parity operators.

\section{Coherent states parametrized by
           $\mathbf{Z_{n}}\times \mathbf{D_{n}}$}

In this section generalized coherent states will be determined for
each of the two quantum kinematics.

{\it A family of generalized coherent states of type $\{\Gamma(g),
\vert\psi_{0}\rangle\}$ in~the sense of Perelomov \cite{Perelomov}
is defined for a~representation $\Gamma(g)$ of a~group $\mathbf{G}$
as a~family of states $\{\vert\psi_{g}\rangle\}$,
$\vert\psi_{g}\rangle=\Gamma(g) \vert\psi_{0}\rangle$, where $g$
runs over the whole group $\mathbf{G}$ and  $\vert\psi_{0}\rangle$
is the `vacuum' vector.}

First take quantum kinematics defined by the system of
imprimitivity $(\mathbf{V_1},\mathbf{E})$. To construct
group--related coherent states of Perelomov type parametrized by
$(a,g) \in \mathbf{Z_{n}}\times \mathbf{D_{n}}$,
we define generalized Weyl operators
\begin{equation}\label{}
 \widehat{\mathbf{W_1}}(a,g)=
      \exp(\frac{2\pi ia}{n}\widehat{\mathbf{Q}})
 exp(-i\widehat{\mathbf{P_1}}_{g})=
           e^{\frac{2\pi ia}{n}\widehat{\mathbf{Q}}}\mathbf{V_{1}}(g);
 \quad a\in \mathbf{Z_{n}},\, g\in \mathbf{D_{n}}.
\end{equation}
Here
\begin{equation}\label{eiaq}
  (e^{\frac{2\pi ia}{n}\widehat{\mathbf{Q}}})_{jk}=
  \delta_{j,k}e^{\frac{2\pi iaj}{n}},\quad
  \exp(\frac{2\pi ia}{n}\widehat{\mathbf{Q}})=
 \left(  \begin{array}{ccccc}
   1 &  &  &  &  \\
    & e^{2\pi \frac{ia}{n}} &  &  &  \\
    &  & \cdot &  &  \\
    &  &  & \cdot &  \\
    & & & & e^{\frac{2\pi ia(n-1)}{n}} \
  \end{array}\right).
\end{equation}
Note that, if the system of imprimitivity is irreducible, also the
set of generalized Weyl operators defined above acts irreducibly in
the Hilbert space $\cal H$.
 Restricting $g$ to the subgroup $\mathbf{Z_{n}}$ of discrete
rotations, the unitary operators satisfy
\begin{equation}\label{xyz}
  e^{\frac{2\pi ia}{n}\widehat{\mathbf{Q}}}e^{im\widehat{\mathbf{P}}}=
  e^{\frac{2\pi iam}{n}}e^{im\widehat{\mathbf{P}}}
  e^{\frac{2\pi ia}{n}\widehat{\mathbf{Q}}}
\end{equation}
and operators $\widehat{\mathbf{W_1}}(a,g)$ form the well--known
projective unitary representation of the group $\mathbf{Z_{n}}\times
\mathbf{Z_{n}}$, which acts irreducibly in the Hilbert space
 ${\cal H} =\mathbb{C}^{n}$ \cite{Weyl,StovTolar}.

Unfortunately, if we want to derive a relation similar to \eref{xyz}
for operators $\widehat{\mathbf{P_1}}_{\mathbf{M}_{k}}$, by
performing the same computation as for $\widehat{\mathbf{P}}$ we
obtain
\begin{equation}\label{prus}
 (e^{\frac{2\pi ia}{n}\widehat{\mathbf{Q}}}
 e^{i\widehat{\mathbf{P_1}}_{\mathbf{M}_{m}}})_{jk}=
 e^{\frac{2\pi ia}{n}(2m-2k)}(e^{i\widehat{\mathbf{P_1}}_{\mathbf{M}_{m}}}
 e^{\frac{2\pi ia}{n}\widehat{\mathbf{Q}}})_{jk}.
\end{equation}
Here the multiplier is $k$--dependent, hence there is neither an
operator equality similar to (\ref{xyz}) nor a projective
representation property of operators $\widehat{\mathbf{W_1}}(a,g)$.

To construct the system of coherent states in $\mathbb{C}^{n}$,
besides the system of operators $\widehat{\mathbf{W_1}}(a,g)$
a properly defined 'vacuum' vector $|0\rangle$ is needed. Then
generalized coherent states of type
$\{\widehat{\mathbf{W_1}}(a,g),|0\rangle\}$ are given by
\begin{equation}\label{dks}
  |a,g\rangle_1 = \widehat{\mathbf{W_1}}(a,g)|0\rangle, \quad a \in
  \mathbf{Z}_{n}, \, g \in \mathbf{D}_{n},
\end{equation}
and $|0\rangle= |0,e\rangle_1$.
In analogy with continuous case where the coherent states are
eigenvectors of the annihilation operator and the vacuum vector
belongs to eigenvalue $0$ one would like to have a similar condition
\cite{TCh}
\begin{equation}\label{en1}
 e^{\frac{2\pi}{n}\widehat{\mathbf{Q}}}
  e^{i\widehat{\mathbf{P}}}|0\rangle=|0\rangle.
\end{equation}
But (\ref{en1}) cannot hold true since $1$ is not an eigenvalue of the
operator. So our admissible vacuum vectors are required to satisfy
(\ref{en1}) up to a non--zero multiplier \cite{TCh},
\begin{equation}\label{en2}
 e^{\frac{2\pi}{n}\widehat{\mathbf{Q}}}
  e^{i\widehat{\mathbf{P}}}|0\rangle= \lambda|0\rangle.
\end{equation}
For $n$ spectral values
\begin{equation}\label{}
  \sigma(e^{\frac{2\pi}{n}\widehat{\mathbf{Q}}}e^{i\widehat{\mathbf{P}}})
    =  \{\lambda_{k}=e^{\frac{\pi (n-1)}{n}}e^{\frac{2\pi ik}{n}}\;
  \vert \; k=0,1,..,n-1\}
\end{equation}
we obtain a system of $n$ admissible (normalized) vacuum vectors
$|0\rangle^{(k)}$ labeled by $k=0,1,..,n-1$,
\begin{equation}\label{}
  |0\rangle^{(k)}=\mathcal{A}_{n} \left(
   \begin{array}{c}
    1 \\
     e^{\frac{\pi(3-n)}{n}}e^{\frac{-2\pi ik}{n}}\\
    \cdot \\
    \cdot \\
    e^{\frac{\pi(n-1)}{n}}e^\frac{-2\pi ik(n-1)}{n}
  \end{array}\right);
\end{equation}
here the $j$--th component
\begin{equation}\label{g}
(|0\rangle^{(k)})_j=g_{j}^{(k)}=  \mathcal{A}_{n}
  e^{\frac{\pi j(j-n+2)}{n}}e^{-j\frac{2\pi ik}{n}},
\end{equation}
where $j=0,1,\ldots,n-1$ and $\mathcal{A}_{n}$ is the normalization
constant
\begin{equation}\label{A}
  \mathcal{A}_{n}=\frac{1}{\sqrt{\sum_{j=0}^{n-1}e^{\frac{2\pi}{n}j(j-n+2)}}}.
\end{equation}

Now we are able to construct $n$ families of coherent states in the
first quantum kinematics which are labeled by parameter $k$.
Applying (\ref{dks}) for $\mathbf{R}_{m}$, we obtain
\begin{eqnarray}\label{vacr}
 (|a,\mathbf{R}_{m}\rangle^{(k)}_{1})_{j} & = &
 (\widehat{\mathbf{W_{1}}}(a,\mathbf{R}_{m})|0\rangle^{(k)})_{j} =\\
 \nonumber = (e^{\frac{2\pi ia}{n}\widehat{\mathbf{Q}}}\widehat{\mathbf{V_{1}}}
 (\mathbf{R}_{m})|0\rangle^{(k)})_{j} & = & e^{\frac{2\pi iaj}{n}}
 g^{(k)}_{j-m \pmod{n}};
\end{eqnarray}
for $\mathbf{M}_{m}$ we obtain
\begin{eqnarray}\label{vacm}
 (|a,\mathbf{M}_{m}\rangle^{(k)}_{1})_{j}& = &
 (\widehat{\mathbf{W_{1}}}(a,\mathbf{M}_{m})|0\rangle^{(k)})_{j} =\\
 \nonumber = (e^{\frac{2\pi ia}{n}\widehat{\mathbf{Q}}}\widehat{\mathbf{V_{1}}}
 (\mathbf{M}_{m})|0\rangle^{(k)})_{j}
& = & e^{\frac{2\pi iaj}{n}}g^{(k)}_{m-j \pmod{n}}.
\end{eqnarray}

Coherent states  for the second quantum mechanics with
representation $\mathbf{V_{2}}$ are equivalent to those of the first
one because they differ on $\mathbf{M}_{m}$ by an unessential
phase factor $-1$:
\begin{equation}
|a,\mathbf{R}_{m}\rangle^{(k)}_{2}=|a,\mathbf{R}_{m}\rangle^{(k)}_{1},
\qquad
|a,\mathbf{M}_{m}\rangle^{(k)}_{2}=-|a,\mathbf{M}_{m}\rangle^{(k)}_{1}.
\end{equation}

\section{Properties of coherent states}

One of the most important properties of coherent states is their
overcompleteness expressed by a resolution of unity
\begin{equation}\label{}
  \sum_{(a,g)\in \mathbf{Z_{n}}\times \mathbf{D_{n}}}
  |a,g\rangle^{(k)} \langle a,g|^{(k)}=c_{k}\widehat{\mathbb{I}},
\end{equation}
where $c_{k}$ is some non--zero complex number. Let us check this
property for our coherent states. From (\ref{vacr}) and (\ref{vacm})
we get
\begin{equation*}
  \sum_{(a,g)\in \mathbf{Z_{n}}\times \mathbf{D_{n}}}
  |a,g\rangle^{(k)}_{1,2}  \langle a,g|^{(k)}_{1,2}
  =\sum_{a\in \mathbf{Z_{n}},m = 0,..,n-1}
  |a,\mathbf{R}_{m}\rangle^{(k)}_{1} \langle a,\mathbf{R}_{m}|^{(k)}_{1}
  \end{equation*}
  \begin{equation}\label{rou}
  + \sum_{a \in \mathbf{Z_{n}},m=0,..,n-1}
  |a,\mathbf{M}_{m}\rangle^{(k)}_{1} \langle a,\mathbf{M}_{m}|^{(k)}_{1}.
\end{equation}
Matrix element of the first sum on the right--hand side of
(\ref{rou}) is, due to (\ref{g}) and (\ref{A}),
\begin{equation*}
 (\sum_{a,m}|a,\mathbf{R}_{m}\rangle^{(k)}_{1} \langle
  a,\mathbf{R}_{m}|^{(k)}_{1})_{jl}
 =\sum_{a,m}(|a,\mathbf{R}_{m})\rangle^{(k)}_{1})_{j} (\langle
 a,\mathbf{R}_{m}|^{(k)}_{1})_{l} = \end{equation*}
 \begin{equation}\label{rour}
  =\sum_{a,m}e^{\frac{2\pi ia}{n}(j-l)}
  g^{(k)}_{j-m \pmod{n}}\overline{g^{(k)}_{l-m \pmod{n}}}
 = n\delta_{j,l}\langle 0|0\rangle^{(k)}=n\delta_{j,l}.
\end{equation}
Exactly the same result is obtained for the second sum on the
right--hand side of (\ref{rou}):
\begin{eqnarray}\label{roum}
 \nonumber (\sum_{a,m}|a,\mathbf{M}_{m}\rangle^{(k)}_{1} \langle
  a,\mathbf{M}_{m}|^{(k)}_{1})_{jl} =\sum_{a,m}e^{\frac{2\pi ia}{n}(j-l)}
  g^{(k)}_{m-j \pmod{n}}\overline{g^{(k)}_{m-l \pmod{n}}} = \\
   = n\delta_{j,l}\sum_{m}g^{(k)}_{m-j \pmod{n}}
   \overline{g^{(k)}_{m-l \pmod{n}}} =n\delta_{j,l}.
\end{eqnarray}
So we have proved that the resolution of unity is fulfilled:
\begin{equation}\label{rouf}
 \sum_{(a,g)\in \mathbf{Z_{n}}\times \mathbf{D_{n}}}|a,g\rangle^{(k)}_{1,2}
\langle a,g|^{(k)}_{1,2}=2n\widehat{\mathbb{I}}
\end{equation}
and this result holds for both representations $\mathbf{V_{1}}$ and
$\mathbf{V_{2}}$.

For the inner product (overlap) of two coherent states we have the
formulae
\begin{eqnarray}\label{sumsum}
 \langle
 a,\mathbf{R}_{p}|b,\mathbf{R}_{q}\rangle^{(k)}_{1,2}
 &=&\sum_{j=1}^{n}e^{\frac{2\pi ij}{n}(b-a)}
 \overline{g^{(k)}_{j-p \pmod{n}}}g^{(k)}_{j-q \pmod{n}},\\
 \nonumber
 \langle a,\mathbf{M}_{p}|b,\mathbf{M}_{q}\rangle^{(k)}_{1,2}
 &=&\sum_{j=1}^{n}e^{\frac{2\pi ij}{n}(b-a)}
 \overline{g^{(k)}_{p-j \pmod{n}}}g^{(k)}_{q-j \pmod{n}},\\
 \nonumber \langle
 a,\mathbf{R}_{p}|b,\mathbf{M}_{q}\rangle^{(k)}_{1,2}
 &=&\sum_{j=1}^{n}e^{\frac{2\pi ij}{n}(b-a)}
 \overline{g^{(k)}_{j-p \pmod{n}}}g^{(k)}_{q-j \pmod{n}}.
\end{eqnarray}
 Note that the inner products yield the reproducing kernel
$ \langle x \vert x' \rangle = K(x,x')$ \cite{AAG}.

If the system is prepared in the coherent state
$|a,g\rangle^{(k)}_{1,2}$, then the probability to measure the
eigenvalue $j$ of position operator is given by $\vert \langle j
|a,g\rangle^{(k)}_{1,2}\vert^2$. It is independent of $k$ and is the
same in both quantum kinematics, namely,
 \begin{eqnarray}
 \vert\langle j|a,\mathbf{R}_{m}\rangle^{(k)}_{1,2}\vert^2 &=&
  \mathcal{A}_{n}^{2} e^{\frac{2\pi}{n}(j-m)(j-m-n+2)}, \cr
 \vert\langle j|a,\mathbf{M}_{m}\rangle^{(k)}_{1,2}\vert^2 &=&
 \mathcal{A}_{n}^{2}e^{\frac{2\pi}{n}(m-j)(m-j-n+2)}.
 \end{eqnarray}

\section{Concluding remarks}
In this paper we have constructed systems of imprimitivity on the
finite configuration space $\mathbf{Z}_{n}$ considered as a
homogeneous space of the dihedral group $\mathbf{D}_{n}$. We have
shown that there exist two inequivalent irreducible systems of
imprimitivity $(\mathbf{V_{1}},\mathbf{E})$ and
$(\mathbf{V_{2}},\mathbf{E})$. Unitary representations
$\mathbf{V_{1}}$ and $\mathbf{V_{2}}$ have clear physical
significance of symmetry transformations.

Using these systems of imprimitivity, we have constructed the
corresponding families of group related coherent states in the sense
of Perelomov. They are connected with the group
$\mathbf{Z}_{n}\times \mathbf{D}_{n}$ acting on the discrete phase
space $\mathbf{Z}_{n}\times \mathbf{Z}_{n}$. Unfortunately, due to
(\ref{prus}) we have lost the group property of the set of operators
$\widehat{\mathbf{W}}(a,g)$, i.e. these operators do not form a
projective unitary representation of the group $\mathbf{Z}_{n}
\times \mathbf{D}_{n}$. In spite of this fact for the first system
of imprimitivity $n$ families of coherent states were obtained,
generated from $n$ admissible vacuum vectors (\ref{g}). It turned
out that the coherent states for the second system of imprimitivity
differ from the first only by an unessential phase factor, i.e.,
they are physically equivalent. For all $n$ families of coherent
states the overcompleteness property was demonstrated. We have also
evaluated the overlaps of pairs of coherent states in the form of
finite sums (\ref{sumsum}). The only physical difference between the
two quantum kinematics can be observed in the difference between
unitary representations $\mathbf{V_{1}}$ and $\mathbf{V_{2}}$ on
mirror symmetries, which have the meaning of parity operators.

Let us note that in quantum optics, discrete phase space
$\mathbf{Z}_{n}\times \mathbf{Z}_{n}$ is employed in connection with
the quantum description of phase conjugated to number operator
\cite{PeggBarnett}. Our approach can also provide a suitable
starting point for the approximate solution of the continuous
Schr\"odinger equation. In this connection we found instructive the
paper \cite{Digernes} on finite approximation of continuous Weyl
systems inspired by an approximation scheme due to J. Schwinger
\cite{Schwinger*}.

Another interesting application is offered by quantum chemistry,
viz. H\"uckel's treatment of delocalized $\pi$-electrons and its
generalizations in various kinds of molecules, where molecular
orbitals are expressed as linear combinations of atomic orbitals
\cite{Rouvray,Ruedenberg}. In this respect our approach seems
especially suitable for the treatment of ring molecules with $n$
equivalent carbon atoms called annulenes . In our notation, the set
of atomic orbitals would correspond to the standard basis in
$\mathcal{H} = \mathbb{C}^n$ and unitary representations
$\mathbf{V_{1}}$ and $\mathbf{V_{2}}$ realize the geometric symmetry
transformations.

\section*{Acknowledgements}
The authors are indebted to A. Odzijewicz and M. Znojil for useful
discussions. J.T. thanks J. Patera and Centre de Recherches
Math\'ematiques, Universit\'e de Montr\'eal for hospitality. Partial
support by the Ministry of Education of Czech Republic (projects
MSM6840770039 and LC06002) is gratefully acknowledged.

\section*{Appendix}
For computation of matrix functions the Lagrange--Sylvester theorem
is useful:

\noindent {\bf Theorem} \cite{LSF}.{\it Let $\mathbb{A}$ be an
$n\times n$ matrix with spectrum $\sigma(\mathbb{A})=\{
\lambda_{1},\lambda_{2},...,\lambda_{s}\}$, $s\leq n$. Let $q_{j}$
be the multiplicity of eigenvalue $\lambda_{j}$, $j=1,2,...,s$. Let
$\Omega \subset \mathbb{C}$ be an open subset of the complex plane
such that $\sigma(\mathbb{A})\subset \Omega$. Then the formula
\begin{equation}\label{LSF}
  f(\mathbb{A}) =
  \sum_{j=1}^{s}\sum_{k=0}^{q_{j}-1}\frac{f^{(k)}(\lambda_{j})}{k!}
  (\mathbb{A}-\lambda_{j}\mathbb{I})^{k}\mathbb{P}_{j}
\end{equation}
holds for every function $f$ holomorphic on $\Omega$. Here
$\mathbb{P}_{j}$ is the orthogonal projector onto the subspace of
$\mathbb{C}^{n}$ which is spanned by the set of all eigenvectors
with eigenvalue $\lambda_{j}$:
\begin{equation}\label{proj}
  \mathbb{P}_{j}:=\prod_{l=1,l\neq
  j}^{s}\frac{\lambda_{l}\mathbb{I}-\mathbb{A}}{\lambda_{l}-\lambda_{j}}.
\end{equation}
}

 The formula (\ref{LSF}) can be applied to equation (\ref{ln}) to
evaluate operators $\widehat{\mathbf{P_1}}_{g}$ for mirror
symmetries. Since the multiplicities of spectral values
$\pm 1$ have already been determined, we
have only to find the spectral projectors $\mathbb{P}_{k}$ for each
representation element $\mathbf{V_{1}}(\mathbf{M}_{k})$.
From equation
(\ref{ln})
\begin{equation}\label{}
 \widehat{\mathbf{P_1}}_{\mathbf{M}_{k}}= i\cdot
 \ln(\mathbf{V_{1}}(\mathbf{M}_{k})),
\end{equation}
we get, using the Lagrange--Sylvester formula (\ref{LSF})
with spectrum (\ref{spm}), the spectral decomposition
\begin{eqnarray}\label{ppp}\nonumber
 \widehat{\mathbf{P_1}}_{\mathbf{M}_{k}}=i\cdot
 \sum_{j=0}^{q_{(+)}-1}\frac{\ln^{(j)}(+1)}{j!}(\mathbf{V_{1}}
 (\mathbf{M}_{k})-\mathbb{I})^{j}\widehat{\mathbb{P}}_{+1}    \\
 +i\cdot \sum_{j=0}^{q_{(-)}-1}\frac{\ln^{(j)}(-1)}{j!}(\mathbf{V_{1}}
 (\mathbf{M}_{k})+\mathbb{I})^{j}\widehat{\mathbb{P}}_{-1},
\end{eqnarray}
where $q_{(\pm)}$ are multiplicities of eigenvalues $\pm 1$.
Strictly said the assumption of the Lagrange--Sylvester formula
(\ref{LSF}) is not satisfied since the complex logarithm is not
holomorphic on the non--positive part of the real axis and $-1$
belongs to the spectrum of $\mathbf{V_{1}}(\mathbf{M}_{k})$. We will
express $\widehat{\mathbf{P}}_{\mathbf{M}_{k}}$  in a formal way and
verify (\ref{exp}) using (\ref{LSF}), where function $\exp$ is
holomorphic.

 Using formula \eref{proj} for the projectors projecting
on  $q_{(\pm)}$-dimensional subspaces of $\mathbb{C}^n$
\begin{equation}\label{}
  \widehat{\mathbb{P}}_{+1}  =
  \frac{(\mathbf{V_{1}}(\mathbf{M}_{k})+\mathbb{I})}{2}, \qquad
\widehat{\mathbb{P}}_{-1} =
 -\frac{(\mathbf{V_{1}}(\mathbf{M}_{k})-\mathbb{I})}{2},
\end{equation}
and the property
\begin{equation}\label{}
  (\mathbf{V_{1}}(\mathbf{M}_{k})-\mathbb{I})
  (\mathbf{V_{1}}(\mathbf{M}_{k})+\mathbb{I})=
  (\mathbf{V_{1}}(\mathbf{M}_{k}))^{2}-\mathbb{I}=\widehat{0},
\end{equation}
all elements in the sum  \eref{ppp} vanish except $j=0$:
\begin{equation}\label{gz}
 \widehat{\mathbf{P_1}}_{\mathbf{M}_{k}} =  i\cdot
 (\frac{\ln(+1)}{2}(\mathbf{V_{1}}(\mathbf{M}_{k})+
 \mathbb{I})-\frac{\ln(-1)}{2}(\mathbf{V_{1}}(\mathbf{M}_{k})-\mathbb{I})).
\end{equation}
Taking the value $-\pi$ for $\ln (-1)$
\begin{equation}
\widehat{\mathbf{P_1}}_{\mathbf{M}_{k}}
= \frac{\pi}{2}(\mathbf{V_{1}}(\mathbf{M}_{k})-\mathbb{I});
\end{equation}
similar calculation leads to
\begin{equation}\label{}
  \widehat{\mathbf{P_2}}_{\mathbf{M}_{k}}=
  \frac{\pi}{2}(\mathbf{V_{2}}(\mathbf{M}_{k})-\mathbb{I}).
\end{equation}
Note that momentum operators are not uniquely determined. This is
caused by the property of exponential mapping which is not
one-to-one.


\section*{References}
\begin{thebibliography}{99}
% By default the references are printed smaller in the iopart style
\normalsize
%\bibitem{fn83}
%Fang M T C and Newland D B 1983 J. Phys. D: Appl. Phys. 16 793-810

%\bibitem{a65}
%Ames W F 1965 Nonlinear Differential Equations in Engineering (New
%York: Academic)

%\bibitem{gm71}
%Gordon S and McBride J 1971 NASA Report SP-273

\bibitem{Weyl}
Weyl H 1931 {\it The Theory of Groups and Quantum Mechanics} (New
York: Dover) pp 272--280

\bibitem{Schwinger}
Schwinger J 1970 {\it Quantum Kinematics and Dynamics} (New York:
Benjamin) pp 63--72

\bibitem{Tolar}
Tolar J 1977 {\it Quantization Methods} lecture notes, Institut
f\"ur Theoretische Physik der Technischen Universit\"at Clausthal

\bibitem{StovTolar}
\v{S}\v{t}ov\'{\i}\v{c}ek P and Tolar J 1984 Quantum mechanics in a
discrete space-time {\it Rep. Math. Phys.} {\bf 20} 157--170

\bibitem {Mackey}
Mackey G W 1968 {\it Induced Representations and Quantum Mechanics}
(New York: Benjamin)

\bibitem{HDDTolar}
Doebner H--D and Tolar J 1975 Quantization on homogeneous spaces
{\it J. Math. Phys.} {\bf 16} 975--985

\bibitem{HDDStovTolar}
Doebner H--D, \v S\v{t}ov\'{\i}\v cek P and Tolar J 2001
Quantization of kinematics on configuration manifolds {\it Rev.
Math. Phys.} {\bf 13} 799--845

\bibitem{Perelomov}
Perelomov A M 1986 {\it Generalized Coherent States and Their
Applications} (Berlin: Springer)

\bibitem{TCh}
Tolar J and Chadzitaskos G 1997 Quantization on $\mathbf{Z}_{M}$ and
coherent states over $\mathbf{Z}_{M}\times\mathbf{Z}_{M}$ {\it J.
Phys. A: Math. Gen.} {\bf 30} 2509--2517

\bibitem{Coleman}
Coleman A J 1968 Induced and subduced representations {\it Group
Theory and its Applications, Vol. 1} ed E M Loebl (New York:
Academic Press)

\bibitem{BEH}
Blank J, Exner P and Havl\'{i}\v{c}ek M 1994 {\it Hilbert--Space
Operators in Quantum Physics} (New York: American Institute of
Physics)

\bibitem{AAG}
Ali S T, Antoine J--P and Gazeau J--P 2000 {\it Coherent States,
Wavelets and Their Generalizations} (New York: Springer)

\bibitem{PeggBarnett}
Pegg D T and Barnett S M 1988 Unitary phase operator in quantum
mechanics {\it Europhys. Lett.} {\bf 6} 483--7

\bibitem{Digernes}
Digernes T, Husstad E and Varadarajan V S 1999 Finite approximation
of Weyl systems {\it Math. Scand.} {\bf 84} 261--283

\bibitem{Schwinger*}
Schwinger J 1960 {\it Proc. Nat. Acad. Sci. U.S.A.} {\bf 46}
570--579, 1401--1415

\bibitem{Rouvray}
Rouvray D H 1976 The topological matrix in quantum chemistry {\it
Chemical Applications of Graph Theory} ed A T Balaban (New York:
Academic Press) 175--221

\bibitem{Ruedenberg}
Ruedenberg K and Scherr C W 1953 Free--electron network model for
conjugated systems I. Theory {\it J. Chem. Phys.} {\bf 21} 1565--81

\bibitem{LSF}
Lancaster P and Tismenetsky M 1985 {\it The Theory of Matrices with
Applications} 2nd ed (New York: Academic Press)

%\bibitem{Klauder}   {\it Coherent States - Applications in Physics
%    and Mathematical Physics}
%  (eds. J.R. Klauder and B.-S. Skagerstam), World Scientific, Singapore 1985.
%\bibitem {Per2} {A.M. Perelomov, On the completeness of a system of coherent
% states, {\it Teor. Mat. Fiz.} {\bf 6}(2) (1971), 213-224}
%\bibitem{Luft} P. Luft, {\it Quantization and coherent states},
% diploma thesis, Czech Technical University, Prague 2003

\end{thebibliography}
\end{document}